\documentclass[useAMS,usenatbib]{mn2e}

\usepackage{mathptmx} 
\usepackage{natbib}
\usepackage[]{graphicx}
\usepackage{xcolor}
\usepackage{dsfont}
\usepackage{bm}
\usepackage{wasysym}
\usepackage{undertilde}
\usepackage{aas_macros_long}
\usepackage{url}
\usepackage{multirow}

\definecolor{midblue}{rgb}{0,.43,1}
\definecolor{aqua}{rgb}{0,1,1}

\newcommand{\undertilde}[1]{\utilde{#1}}

\newcommand{\executeiffilenewer}[3]{%
 \ifnum\pdfstrcmp{\pdffilemoddate{#1}}%
 {\pdffilemoddate{#2}}>0%
 {\immediate\write18{#3}}\fi%
}

\title[Stellar Imaging with \emph{Cassini} Occultations II]{High Angular Resolution Stellar Imaging with Occultations from the \emph{Cassini} Spacecraft II: Kronocyclic Tomography}

\author[Paul N. Stewart et al.]
{\parbox{\textwidth}{Paul N. Stewart$^{1}$\thanks{E-mail:p.stewart@physics.usyd.
edu.au (PNS)},
Peter G. Tuthill$^{1}$,
Philip D. Nicholson$^{2}$,
Matthew M. Hedman$^{3}$,
and James P. Lloyd$^{2}$}\vspace{0.4cm}\\
$^{1}$Sydney Institute for Astronomy, School of Physics, H90, The University of 
Sydney, NSW 2006, Australia\\
$^{2}$Department of Astronomy, Cornell University, Ithaca, NY 14853, USA\\
$^{3}$Physics Department, University of Idaho, Moscow, ID 83844, USA
}

\begin{document}

\date{Accepted . Received ; in original form }

\pagerange{\pageref{firstpage}--\pageref{lastpage}} \pubyear{2002}

\maketitle

\label{firstpage}

\begin{abstract}
We present an advance in the use of \emph{Cassini} observations of stellar occultations by the rings of Saturn for stellar studies.
\citet{Stewart2013} demonstrated the potential use of such observations for measuring stellar angular diameters.
Here, we use these same observations, and tomographic imaging reconstruction techniques, to produce two dimensional images of complex stellar systems.
We detail the determination of the basic observational reference frame.
A technique for recovering model-independent brightness profiles for data from each occulting edge is discussed, along with the tomographic combination of these profiles to build an image of the source star.
Finally we demonstrate the technique with recovered images of the $\alpha$ Centauri binary system and the circumstellar environment of the evolved late-type giant star, Mira.
\end{abstract}


\section{Introduction}
\label{intro}

Often, a great deal of useful information about celestial targets may be obtained from sparse imaging data, such as one-dimensional brightness profiles.
A novel example of such data was presented in~\citet{Stewart2013} (Paper I) which demonstrated the use of \emph{Cassini}-VIMS 
(Visible and Infrared Mapping Spectrometer) observations of stellar occultations by the Saturnian rings to recover high spatial resolution information about the circumstellar environment of Mira.

However, there are circumstances in commonly encountered astronomical systems in which simple assumptions such as spherical symmetry of a target object are no longer reliable.
In some of the most extreme examples, such as imaging the close environments of very dusty mass-losing evolved stars \citep{Tuthill2005}, the recovered morphology may have very few residual symmetries or predictable structural elements with which to build models for low-order parameter fitting. 
In such cases, the only recourse is some form of model-independent image recovery process.

Although Paper~1 primarily employed the \emph{Cassini} occultation data to recover a simple one-dimensional brightness profile, the data themselves do explore more of the object's flux distribution than this.
In particular, over the course of single passage of the entire Saturnian ring system across the sightline to a celestial target, we found that there were usually a handful of usable occultation events at various known sharp edges within the ring structure (see Paper~1 for an explanation of the basic occultation geometry).
Most importantly, the apparent sky-orientation of the occulting edge for each occultation event is a different angle, so that an ensemble of such events -- built up over the course of one or several complete passes behind the rings -- yields a data set capable of constraining a full two dimensional image.

More generally the process of reconstructing an image from a series of overlapping one-dimensional projections is known as Tomography.
This has been particularly important in medical imaging for the past several decades as a non invasive way to view structures internal to the human body.
It turns out that the data recovered from \emph{Cassini}-VIMS occultations form a sufficiently similar representation that we are able to employ these same techniques to derive two dimensional images of stellar sources and their circumstellar environments.

The first problem confronted in the image formation process is recovery of the angle on-sky from each occulting edge in a suitable celestial reference frame.
This is discussed in Section~\ref{sec:find_PA}.
The process of generating tomographic reconstructions of stellar targets themselves is detailed in Section~\ref{sec:tomo}, and is broken into steps.
The first (Section~\ref{sec:MIBPR}) is to recover the one-dimensional brightness profile of the source star from each occultation observation.
These are then combined tomographically to build an image (Section~\ref{sec:tomo_r}) in an iterative process to determine the projection registration (Section~\ref{sec:cent}).

\section{Position Angle of Projections}\label{sec:find_PA}

In order to make sense of measured asymmetries in fitted brightness profiles, and to register structure against published literature, the projection angle of each one-dimensional profile must be determined.
This is the angle normal to the occulting edge as viewed in the sky-plane of \emph{Cassini}, relative to north in the celestial sphere of the Earth. 
The spacecraft position and pointing direction are recorded by mission controllers in a kronocentric reference frame, so the terrestrial celestial north vector must be determined relative to that of Saturn's north.

\begin{figure} 
\centering
  \def\svgwidth{.99\columnwidth}
 \executeiffilenewer{PA_geom2.svg}{PA_geom2.pdf}%
 {inkscape -z -D --file=PA_geom2.svg %
 --export-pdf=PA_geom2.pdf --export-latex}%
 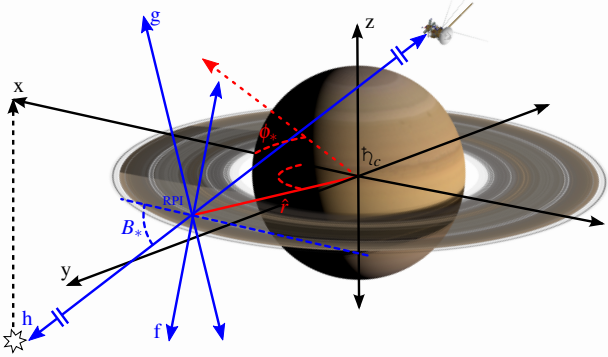%

  \caption{An example of the two coordinate systems used to compare vectors from different sources.
  The blue vectors identify the $f$,$g$,$h$ coordinate system, with $h$ along the line of sight from Cassini to the star, and centred on the point where this line intercepts the ring plane, the RPI (Ring Plane Intercept).
  The $f$,$g$ coordinates define Cassini's sky-plane with $f$ lying in the ring plane.
  The black vectors show the $x$,$y$,$z$ coordinate system with its origin at the centre of the planet, $z$ through the poles and $x$ defined by the ring plane projection of the vector towards the star.
  The red line ($\hat{r}$) joins the origins of the two coordinate systems and $\phi$ is the angle between the $x$ axis and the kronocentric longitude of the RPI.
  The red dashed vector is the standard kronocentric $x$ direction in the ring plane, and $\phi_*$ is the angle between this and the kronocentric longitude of the star.
  The projection of the $h$ axis into the ring plane is parallel to the $x$ axis, and the angle between the $h$ axis and this projection is $B_*$.}
  \label{fig:PA_geom2} 
\end{figure}

We start by defining two coordinate systems: a kronocentric reference frame with its origin at the centre of Saturn, and another centred on the point where the planetary ring plane intercepts the line of sight from \emph{Cassini} to the star as shown in Figure~\ref{fig:PA_geom2}.
Using these two coordinate systems simplifies the process of determining the required angle using vectors coming from different sources.

The kronocentric coordinate system has its origin at Saturn's centre ($\saturn_c$) with $z$ through the planet's poles and $x$ and $y$ in the ring plane.
The standard $0^\circ$ longitude in the kronocentric coordinate frame is defined to be the ascending node of the ring-plane on the terrestrial equatorial plane in J2000.
In order to simplify the geometry, we instead define the $x$ axis to be the ring-plane projection of the vector from $\saturn_c$ toward the star.
This direction is effectively fixed for any particular star, varying only with planetary precession and stellar proper motion on million year time-scales.
This is achieved by rotating by the kronian longitude of the star ($\phi_*$) around the z axis,

\begin{equation}\label{eq:R3}
 \quad [x,y,z]=\mathds{R}_3(\phi_*)[x,y,z]
\end{equation}

\begin{figure} 
\centering
  \def\svgwidth{.99\columnwidth}
 \executeiffilenewer{PA_geom.svg}{PA_geom.pdf}%
 {inkscape -z -D --file=PA_geom.svg %
 --export-pdf=PA_geom.pdf --export-latex}%
 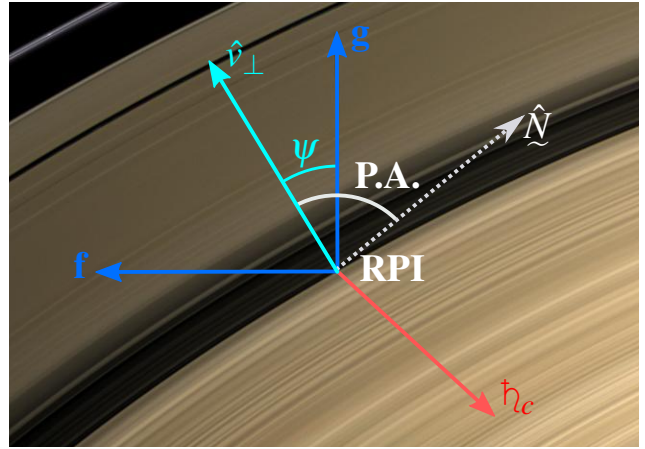%

  \caption{Identifying the important vectors in \emph{Cassini}'s sky-plane (defined by the $f$ and $g$ axes).
  $g$ is the projection of Saturn's north vector translated to the RPI, where the line of sight to the star (unlabelled $h$ axis) is intercepted by the planetary ring-plane. 
  $\saturn_c$ is the direction towards the centre of Saturn.
  $\hat v_\perp$ is the direction normal to the rings which corresponds to the direction of spatial information encoded in the occultation
  ${\protect\undertilde{\hat{N}}}$ is the terrestrial celestial north vector translated to the $RPI$ and $P.A.$ is the position angle of $\hat v_\perp$ relative to the Earth's north.
  Base image:~PIA12518~\citep{CIT2010}}
  \label{fig:PA_geom} 
\end{figure}

The second coordinate system has its origin at the ring-plane intercept (RPI) which is the point where the planetary ring plane is intercepted by the line of sight from \emph{Cassini} to the star, with one axis ($h$) running along the line of sight to the star, and the orthogonal axes ($f$ and $g$) in \emph{Cassini}'s sky-plane.
This coordinate system and the relevant vectors are shown in Figure~\ref{fig:PA_geom}.

The sky-plane coordinates can be defined in terms of the Kronocentric coordinates by

\begin{equation}
 \quad f=y-y_{r}
\end{equation}
\begin{equation}
 \quad g = z \, \cos(B_{*})-(x-x_r)\,\sin(B_{*})
\end{equation}

where $x_r$ and $y_r$ are the x, y coordinates of the RPI and $B_*$ is the kronocentric latitude of the star which is constant for any particular star.
This is effectively just a rotation of $-B_*$ around the $y$ axis as 

\begin{equation}\label{eq:R2}
 \quad [h,f,g] = \mathds{R}_2(-B_*)[x-x_r, y-y_r, z]
\end{equation}

The determination of the radius ($\hat{r}$) and longitude ($\phi$) of the of the RPI is discussed in \citet{Stewart2013}.
From these, the determination of $x_r$ and $y_r$ is trivial.

\begin{equation}
 \quad \hat{r} = (r\cos(\phi), r\sin(\phi),0) \equiv (x,y,z)_r
\end{equation}
 
We are now able to rotate the terrestrial celestial north vector into \emph{Cassini}'s sky-plane by employing three matrix rotations.
Two of the three matrices required are taken from Equation~\ref{eq:R3} and Equation~\ref{eq:R2}, while the final matrix is $\mathds{C}_EP(\alpha_p, \delta_p)$ which rotates J2000 coordinates into kronocentric equatorial coordinates.
These transformations may be combined into a single matrix operation:

\begin{equation}
 \quad \undertilde{\hat{N}} \equiv [h,f,g]_N = \mathds{R}_2(-B_*)\mathds{R}_3(\phi_*)\mathds{C}_EP(\alpha_p, \delta_p)[0,0,1]
\end{equation}

enabling translation of the Earth's celestial north into the RPI/sky-plane coordinate system.

The position angle ($P.A.$) of the occultation can be determined as follows,

\begin{equation}
 \quad P.A. = \frac{cos^{-1}(v_\perp \cdot \undertilde{\hat{N}})}{v_\perp}
\end{equation}

where $v_\perp$ is defined to be the direction normal to the projection of the ring edge into the sky plane and is different for each occulted edge.
This can be found by 

\begin{equation}\label{eq:vperp}
\quad \hat v_\perp = [0,\sin(\psi),\cos(\psi)]\equiv[h,f,g]_{v_{\perp}}
\end{equation}

where $\psi$ is determined from

\begin{equation}\label{eq:tanpsi}
\quad  \tan(\psi) = -\frac{dg}{df}
\end{equation}

The movement of the line of sight to the star in the ($f$,$g$) plane is 

\begin{equation}\label{eq:f}
\quad  f = -\Delta r\sin(\phi)+\Delta \phi r\cos(\phi)
\end{equation}
\begin{equation}\label{eq:g}
\quad  g = [\Delta r\cos(\phi)+\Delta \phi r\sin(\phi)]\sin(B_*)
\end{equation}

with $\Delta r$ and $\Delta \phi$ being the change in position of the RPI.
$\Delta r$ is locally constant whilst $\Delta \phi$ varies across the ring edge.
Differentiating Equations~\ref{eq:f} and~\ref{eq:g} with respect to $r$ and combining into Equation~\ref{eq:tanpsi} produces

\begin{equation}
\quad  \tan(\psi) = -\frac{dg}{df} = -\tan(\phi)\sin(B_*)
\end{equation}

which can be used to solve Equation~\ref{eq:vperp} to produce $v_\perp$ for each occulting edge.

This set of transformations results in the $h$ axis lying in direction of the line of sight to the star, and so it may be ignored here.
With $v_\perp$ and $\undertilde{\hat{N}}$ now rotated to lie in the sky-plane ($f$,$g$), obtaining the angle between them is straightforward.
The convention adopted is that the angle will be $0^\circ$ if they happen to align; it will be positive if $\hat v_\perp$ is to the left of $\undertilde{\hat{N}}$ toward the east in the sky, otherwise negative (toward the west).

\subsection{Angular Diversity of Observations}

\begin{figure} 
\centering
  \def\svgwidth{.99\columnwidth}
 \executeiffilenewer{PA_div.svg}{PA_div.pdf}%
 {inkscape -z -D --file=PA_div.svg %
 --export-pdf=PA_div.pdf --export-latex}%
 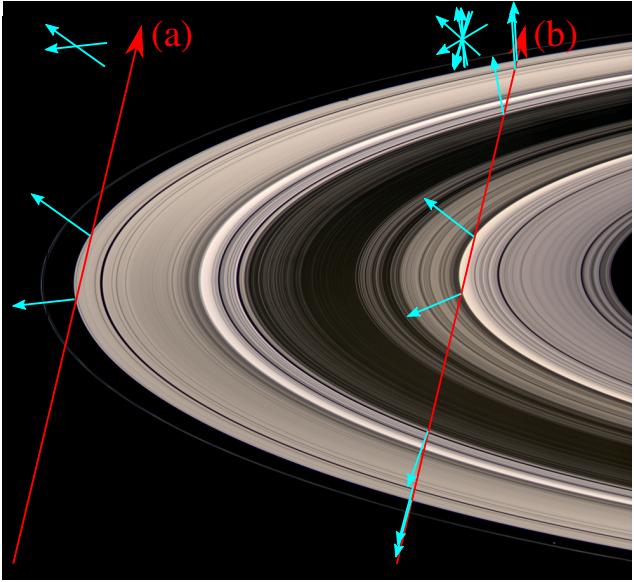%

  \caption{A comparison of angular diversity for representative occultation geometries using indicative edges.
  The long red vectors indicate the apparent trajectory of the star behind the rings.
  The short cyan vectors identify the normal to some of the occulting edges.
  (a) shows a minimal occultation with only two edges that have almost parallel $P.A.$s.
  (b) shows an almost identical orbital geometry, with vastly different angular diversity, although most edges produce almost redundant $P.A.$s.
  The cyan vector clusters to the top left of each shows an angular diversity rose, a quick visual method introduced here to compare angular diversity between observations.
  Base image:~PIA10446~\citep{CIT2010}}
  \label{fig:PA_div} 
\end{figure}

As the star is observed to pass behind the rings, it encounters many edges.
Each of these edges yields an occultation lightcurve from which a brightness profile can be obtained with a different $P.A.$, thus sampling the star in different directions.
Every transit behind the ring plane will produce a different number and diversity of angles.
Figure~\ref{fig:PA_div} shows an example of the differences in angular diversity between two occultations with similar orbital geometry.
The occultation on the left, labelled (a), crosses only two sharp edges, and produces an angular diversity rose with poor coverage, and almost no sampling in the N-S direction.
The vector direction in the ``rose'' is the direction of spatial sampling of the stellar image which is orthogonal to the projection direction.
Occultation (b) has much greater angular diversity, although many of the angles are largely redundant.
It is has far greater sampling in the N-S direction than it has E-W, and is generally superior to occultation (a) in almost all respects for image recovery.

\section{Tomographic Image Reconstruction}\label{sec:tomo}

\subsection{Model-Independent Brightness Profile Recovery}
\label{sec:MIBPR}

Occultations typically recover one-dimensional brightness profiles.
These are the projection of the two-dimensional image of the star in the sky-plane onto a particular axis.
Paper~I demonstrated the successful recovery of circularly symmetric model brightness profiles from \emph{Cassini} occultation observations.
The use of models is a particularly potent way to exploit data which are sparse, noisy or both. 
Best-fit models are able to contribute significantly to studies of target structure, but they are only capable of parameterising structural elements that are already foreseen in the model construction.

On the other hand, the present study attempts to obtain a ``true'' tomographic image, without enforcing symmetries and low-order representation that invariably come with the fitting of models.
As a first step, we need to recover one-dimensional brightness profiles for each occulting edge, independent of any model.
The ideal way to perform this would be to deconvolve the observed lightcurve with the point source lightcurve.
Unfortunately the deconvolution process produces non-unique solutions and is severely under-constrained due to the presence of even small amounts of noise found in all measurements.
For this reason, instead of performing a relatively straightforward Fourier deconvolution, more sophisticated algorithms are required to recover model-independent brightness profiles.

\begin{figure}
 \centering
 \includegraphics[width=.95\columnwidth]{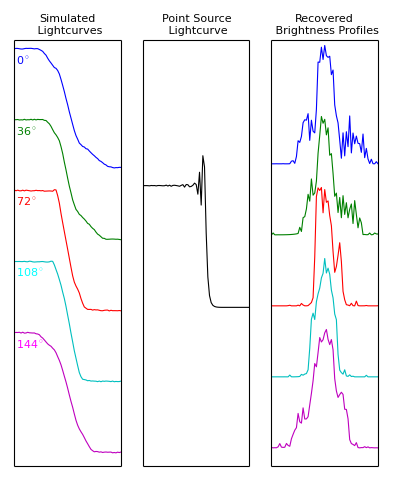}
 \caption{Examples of the model-independent recovery of brightness profiles (right), from simulated lightcurves (left) and the corresponding point source lightcurve (centre).
 The recovered profiles are used in Figure~\ref{fig:centroid}, and the simulated lightcurves are produced from the 'Original Image' in that figure, in the directions indicated by the angular diversity rose.
 The position angle of each lightcurve is labelled on the left hand side in the corresponding colour.
 The occultation geometry used for each lightcurve simulation is identical, resulting in a common point source lightcurve.
 }
 \label{fig:BP}
\end{figure}

The approach taken is to allow a model brightness profile to include a free parameter for each data point in the observation.
Each point is allowed to vary between zero and one before being convolved with the point source lightcurve to produce a model lightcurve.
The point source lightcurve is determined using Fresnel diffraction and the known orbital geometry of Cassini as detailed in \citet{Stewart2013}.
These model lightcurves are in turn compared against the observed lightcurve.
The process iterates by allowing the synthetic brightness profile to evolve towards a better fit.
This ultimately produces a brightness profile which generates a lightcurve closely matching that which has been observed.
Some examples of this process are shown in Figure~\ref{fig:BP}.

The model fitting was performed with the \textit{minimize} function in \textit{Scipy's} \textit{optimize} module.
This is a series of maximum likelihood fitting algorithms which are a native part of the \textit{Scipy} scientific computing environment for the \textit{Python} language.
The ability to regenerate accurate brightness profiles using this method was exhaustively tested by generating synthetic brightness profiles, adding noise, and fitting.
At a signal-to-noise ratio of 100 we are able to recover the overall structure of the brightness profile.
When the signal-to-noise ratio is reduced to 10, we are able to recover the envelope of the brightness profile, but not finer details.

\subsubsection{The Point Source Case}

Ideally an experimental test of the instrumental transfer function would be performed in order to validate the model used to translate observational data into measurements of target morphology; in particular to confirm that departures from idealised representations of the spacecraft detection system and ring edges do not adversely affect the data.
Often, observations of a known unresolved point source reference star are taken to serve this purpose. 
However, the only stars bright enough to be observed with VIMS in occultation mode, are inherently large enough to be resolved, making a true point source observation impossible.
This is a problem sometimes faced by long-baseline ground-based interferometers, where it can be difficult to identify a calibrator with sufficient surface brightness to be detected whilst remaining unresolved.
In the absence of a true point-source, the most rigorous test available is a comparison of fitted model parameters to literature values over a wide spectral range.
This was performed by comparing uniform disc fits to terrestrial interferometric observations of Mira and was presented in~\citet{Stewart2013}.
Close agreement with prior literature was shown to validate results produced from occultations.

\subsection{Back Projection and Regression}
\label{sec:tomo_r}

The building of a tomographic image requires two steps.
These are back projection and regression.
Back projection is projecting the recovered brightness profiles over each other in the direction determined by the $P.A.$
The overlapping profiles immediately form a crude image of the object.
More projections and increased angular diversity provides a visibly better image at this stage, but a limited number of projections or poor angular diversity, as always occurs with these observations, makes the regression step crucial.
Regression allows the image to be modified in some way, such as smoothing or concentrating, whilst maintaining the fit to the back projections.

Tomography is a mature field and has many existing algorithms and tools capable of recovering images from well sampled data.
Most of these tools are not suited to the sparse angular diversity these \emph{Cassini} observations produce.
The selection and implementation of a suitable regression algorithm is therefore critical.
The \textit{Scikit-learn} project~\citep{Pedregosa2011} offers machine learning regression algorithms in \textit{Python} which proved to be sufficiently versatile and capable for image reconstruction.
Their implementation of an elastic net linear regularised regression algorithm~\citep{Zou2005} 
was found to be highly effective with the sparse data produced with this technique.
The selection of this regulariser was made after extensive testing of reconstructions of synthetic objects and found visually to provide the fewest artefacts and the best defined real structure.

\subsection{Centroiding Brightness Profiles}
\label{sec:cent}

\begin{figure}
 \centering
 \includegraphics[width=.75\columnwidth]{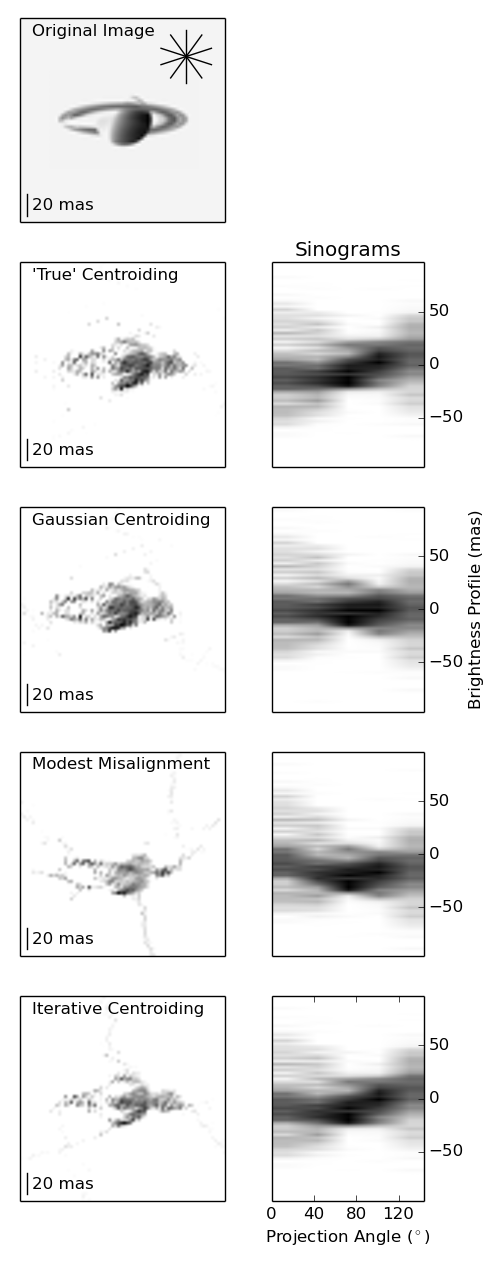}
 \caption{Simulated images demonstrating the impact of centroiding on image recovery.
 The top frame shows the original model image that was used to generate one-dimensional projections.
 The angular diversity rose in the top right of that frame shows the five projection directions.
 These projections were convolved with a point source light curve produced from a typical occultation geometry as shown in Figure~\ref{fig:BP}.
 Realistic noise was added, which remains visible in the reconstructed images as high frequency fluctuations.
 These simulated lightcurves were then fitted with the model-independent process detailed in Section~\ref{sec:MIBPR} to produce brightness profiles.
 The brightness profiles recovered for each projection angle can be seen in the sinograms down the right hand side with corresponding recovered images to their left.
 The vertical axis on the sinograms is at the same scale as the recovered images.
 }
 \label{fig:centroid}
\end{figure}

The brightness profiles as recovered by the algorithms discussed in Section~\ref{sec:MIBPR} possess no unambiguous phase reference, with the exception of simple stellar geometries such as binary systems, or simple circularly symmetric stellar targets.
That is to say that when attempting to back-project more than two brightness profiles taken at different angles to synthesise a two-dimensional image, the phase registration or shift of each projection is not fixed, a scenario not commonly encountered in the tomography literature.
For more complex stellar environments the image reconstruction requires some sort of referencing between projections, otherwise severe artefacts can be introduced in the resultant image.
If the object being imaged is known to be inversion symmetric, this process is quite straightforward and can be accomplished in several different ways with relative ease.
These include centroiding based on fitting Gaussian functions, or registration based on the centre of mass of each of the profiles.

For complex objects known to be dominated by significant asymmetries, or those with unknown geometries, this process can result in misfits and and introduction of artefacts.
The exact registration of an ensemble of angular-diverse brightness profiles cannot be unambiguously determined for an unknown object.
In order to assess the impact of registration on the quality of the output images, a study simulating the entire process from occultation lightcurve to image was performed.
 
Results from these simulations are given in Figure~\ref{fig:centroid}, which depicts the impact upon the final image reconstruction of differing levels of centroiding accuracy.
In this case, a small picture of Saturn (top) was adopted as a useful illustrative model image as it exhibits a distinct asymmetric geometry, with both coarse and fine features ideally suited to benchmarking image recovery techniques.
The ``True'' centroiding aligns the profiles based on exact foreknowledge of the image (and is therefore only possible in a simulation), and shows the upper bound of expected image quality with perfect registration.
The Gaussian centroiding process fits a Gaussian function to the recovered brightness profile and performs a registration based on the function centre.
For this particular case (and any significantly asymmetric object), this approach is not optimal as the brightest part of the image does not occur at the centre and as a consequence the image exhibits artefacts.
For example the ring structure recovered is now asymmetrically disposed with respect to the planet's disc.
The third image/sinogram pair demonstrates that even modest misalignment of the registration can produce artefacts (for example, loss of detail like the elliptical shape of the rings).

After trying several strategies, the most effective method for registering test images was determined to be an iterative process using recovered tomographic images as discussed in Section~\ref{sec:tomo_r}.
At the first iteration, Gaussian centroiding (as above) delivered a trial recovered image.
Next a filtered image was produced by application of a two-pixel Gaussian blur (removing fine detail), and a binary mask applied at the 15\% cut level truncated low level peripheral noise and artefacts.
At the fitting step, the registration of each recovered one-dimensional profile was treated as a free parameter used to minimise the misfit with the filtered image.
Based on this new ensemble of registrations, a new tomographic image was recovered and the algorithm had progressed a full cycle.
This whole process was iterated until none of the profiles were able to provide a better fit with a change in offset.
Given the existence of a known `truth' image in our simulations, we were able to show that the algorithm converged to near the known correct registrations in only a few iterations.
The results of this process are shown in the bottom row of Figure~\ref{fig:centroid}, where the overall shape and balance of the image is recovered with minimal artefacts.

\section{Results of Kronocyclic Tomography}

\subsection{Binary Systems: $\alpha$ Centauri}

The $\alpha$ Centauri system is the closest stellar system to our solar system.
It appears to the eye as the third brightest star in the sky, and the binary nature of $\alpha$ Centauri AB has been observed for over three hundred years.
The orbital elements of the system have been well defined as it is easily resolved with a small telescope and has a long history of measurements.
Modern techniques have been able to substantially refine this, providing highly accurate position predictions.

\begin{figure}
 \centering
 \includegraphics[width=.99\columnwidth]{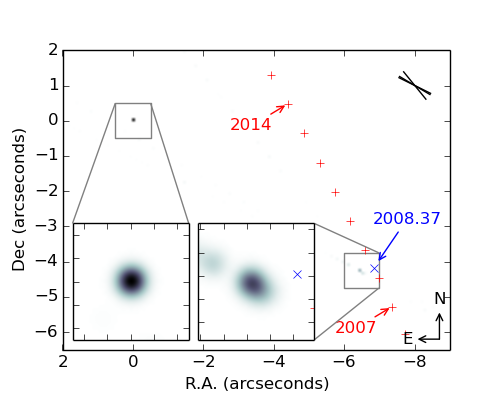}
  \caption{A tomographic reconstruction of the $\alpha$ Cen system from observations performed on 29 April 2008 during \emph{Cassini's} revolution 066.
 The axes are centred on $\alpha$ Cen A, and expressed in arcseconds of right ascension and declination, with north up, and east to the left.
 The red pluses indicate the relative position of $\alpha$ Cen B at the start of each calendar year based on an ephemeris from~\citet{Pourbaix2002}.
 The blue cross, labelled with the decimal year, shows the predicted relative position of $\alpha$ Cen B at this epoch.
 The insets show a 1$''$ zoomed view of each of the stars and the tomographic artefacts in their vicinity.
 }
 \label{fig:tomo_AlpCen066}
\end{figure}

\begin{figure}
 \centering
 \includegraphics[width=.99\columnwidth]{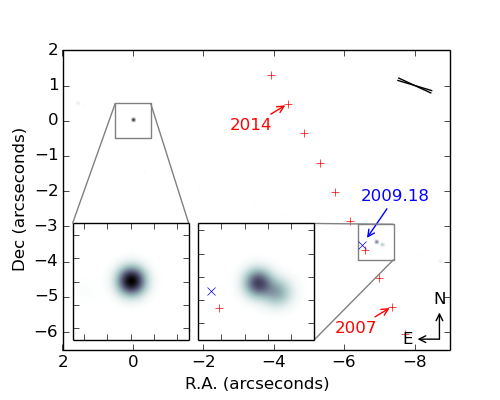}
  \caption{A tomographic reconstruction of the $\alpha$ Cen system from data taken on 6 March 2009 during \emph{Cassini's} revolution 105.
 All details as per Figure~\ref{fig:tomo_AlpCen066}.}
 \label{fig:tomo_AlpCen105}
\end{figure}

The effectiveness of tomographic imaging from \emph{Cassini} stellar occultations was tested by comparing reconstructions from observations of $\alpha$ Centauri at two epochs to the orbital predictions of \citet{Pourbaix2002}.
The first of these reconstructions is shown in Figure~\ref{fig:tomo_AlpCen066}, from an occultation observed on 29 April 2008 (\emph{Cassini's} revolution 066).
The second is from 6 March 2009 (\emph{Cassini's} revolution 105) and is shown in Figure~\ref{fig:tomo_AlpCen105}.
These reconstructions found the pair of stars to be located as expected to within 5\% in separation, and 2$^\circ$ in position angle as shown in Table~\ref{tab:alpCen_table}.

The $\alpha$ Centauri image reconstructions are particularly hindered by the limited angular diversity provided by the specific occultation geometry.
Both epochs suffered from a lack of diversity amongst the projection angles, limiting the ability of the technique to accurately reconstruct the correct position angle.
For R.\,066 the projections were all within a 25$^\circ$ angle, and for R.\,105 they were within 8.1$^\circ$ as shown by the angular diversity roses in Figures~\ref{fig:tomo_AlpCen066} and~\ref{fig:tomo_AlpCen105}.
Both cases also had all their projection angles within 19$^\circ$ of the binary position angle, increasing the uncertainty on the separation.
This would be improved if the position angle and projection angles were closer to perpendicular.
We expect that improved angular diversity would provide substantially closer agreement with the predicted binary separation and position angle.
In spite of these limitations, our observations at these two epochs reveal the expected orbital progression of the binary system.

\begin{table}
\centering
\begin{tabular}{c c c c c}
 Epoch	& \multicolumn{2}{c}{Separation (\phantom{.}$''$)} & \multicolumn{2}{c}{$P.A.$ (\phantom{.}$^\circ$)}\\
 (Year) & Predicted		& Measured		& Predicted		& Measured	\\
 2008.37& 8.007$\pm$0.007		& 7.71$\pm$0.83		& 238.55$\pm$0.09		& 236.5$\pm$2.9\\
 2009.18& 7.399$\pm$0.007		& 7.79$\pm$0.55		& 241.60$\pm$0.11		& 243.6$\pm$2.4\\
\end{tabular} 
\caption{Predictions and results of binary parameters from Alpha Centauri comparing the \citet{Pourbaix2002} ephemeris and tomographic images.}
\label{tab:alpCen_table}
\end{table}

\subsection{Evolved Stars: $\omicron$ Ceti}

$\omicron$ Ceti is the archetype of the Mira Variable stellar class.
These are evolved stars which have exhausted both the hydrogen and helium in their cores, and are known to exhibit complex circumstellar structure~\citep{Tuthill2005}.
Mira itself has been shown to have an asymmetric circumstellar environment~\citep{Karovska1997a, Martin2007} including dusty concentric shell-like structures~\citep{Lopez1997, Cruzalebes1998}.

\begin{figure*}
 \centering
 \includegraphics[width=1.9\columnwidth]{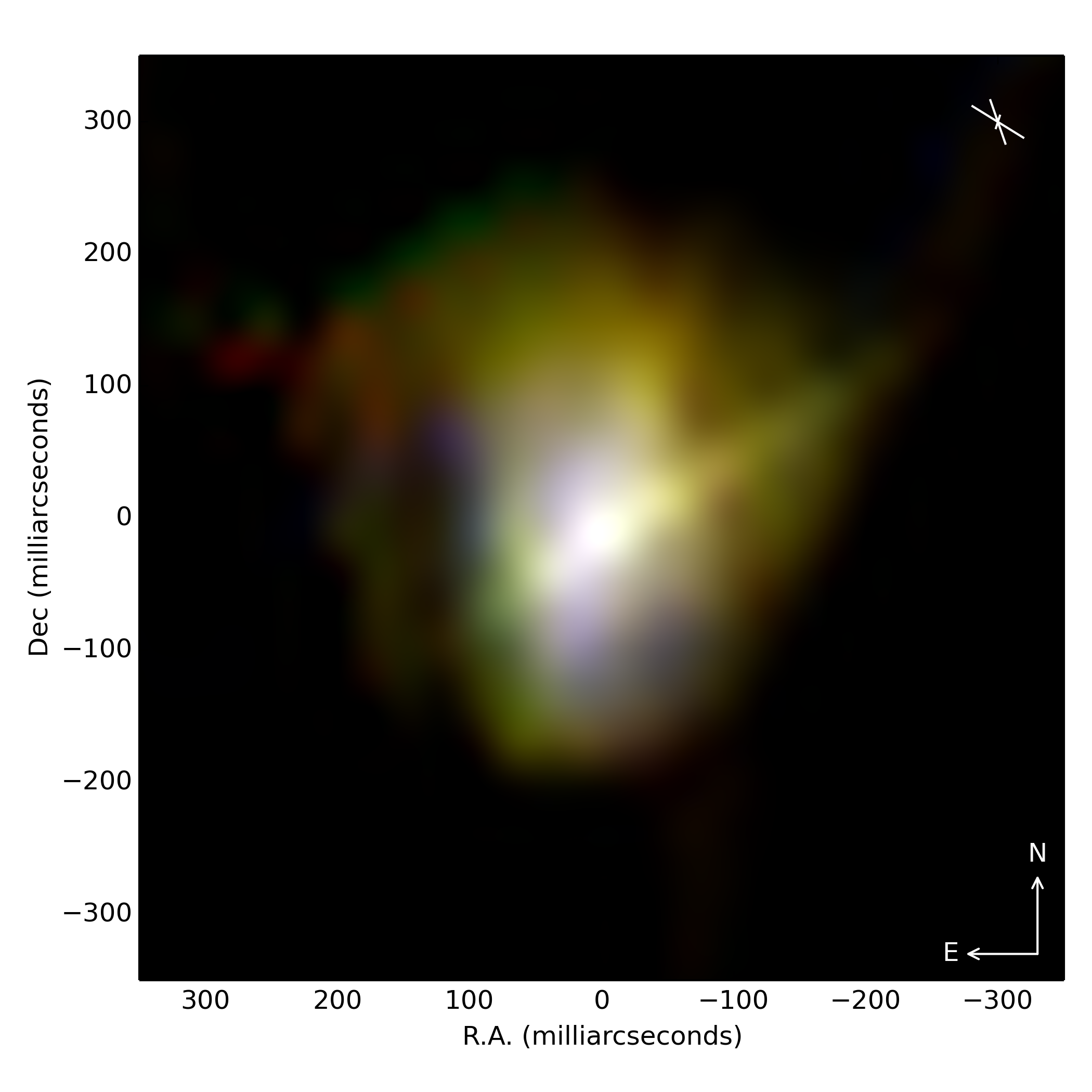}
  \caption{A three colour tomographic reconstruction of Mira from 29 June 2005 during \emph{Cassini's} revolution 10.
  Red is 4.66\,$\mu m$, green is 3.32\,$\mu m$, and blue is 1.34\,$\mu m$.
  North is up and east is left and the axes are in milliarcseconds of right ascension and declination.
  The angular diversity rose in the top right of the image shows the projection directions used in the reconstruction.
  The sampling resolution of each projection is indicated by the length of the corresponding stroke in the rose.
  }
 \label{fig:tomo_Mira010}
\end{figure*}

\citet{Stewart2013} demonstrated the use of \emph{Cassini} stellar occultations to measure the wavelength dependant mean diameter of Mira.
Here we use the same observations from \emph{Cassini's} revolution 10 to tomographically reconstruct a high resolution image of Mira.
Observations from earlier epochs (Revs. 8 and 9) were found to each have relatively poor angular diversity.
The inherent variable nature of Mira prevents using brightness profiles from different epochs being used together to improve this diversity.
The results of this reconstruction are shown in Figure~\ref{fig:tomo_Mira010} revealing the complex structure of the circumstellar environment of Mira.
Occultations by the outer edge of the Keeler Gap on ingress, and the Encke Gap on both ingress and egress were used for this reconstruction.
These particular edges were found to produce a consistent reconstruction with minimal noise.
It shows a bright core with a diameter of around 30\,mas as shown in~\citet{Stewart2013} which contain the vast majority of the stellar flux.
The remaining flux forms a complex dusty environment cooling in discrete steps as a function of radius.
The first of these occurs at a radius of around 80\,mas and the second at a radius of around 200\,mas.
These are consistent with previously identified dynamic H$_2$O shell structures~\citep{Matsuura2002}.
All of these structures are revealed to be somewhat asymmetric with their respective radii varying with angle.
The most obvious of these asymmetries is the enlargement of the outer cooler layer to the north-east of the star.
Similar asymmetries within the inner 100\,mas have previously been identified in the visible part of the spectrum~\citep{Wilson1992, Karovska1997a}.
Figure~\ref{fig:tomo_Mira010} also shows examples of some of the artefacts which this process can introduce.
One of these is the clumpiness of the dusty areas on approximately the same scale as the angular sampling.
Another is the rays which can be seen stretching across the image perpendicular to the sampling directions indicated in the angular diversity rose.
These are a by-product of back projection, and are reduced with improved angular diversity.
The consequence of these artefacts is that fine structure in the resulting images should be treated with scepticism, however the overall envelope appears to be robust.
A detailed astrophysical interpretation, including polychromatic reconstructed images, will be presented in a forthcoming paper.

\section{Conclusions}\label{sec:conc}

We have successfully demonstrated a technique to use stellar occultations by the rings of Saturn, as observed by \emph{Cassini}, to produce tomographic images of stellar systems.
This required a thorough understanding of the spacecraft's orbital geometry, and a translation into a terrestrial reference frame.
It also entailed the extraction of model-independent one dimensional brightness profiles and developing an understanding of tomographic processes.
The process was tested and verified by imaging the known geometry of the $\alpha$ Centauri binary system.
It was then applied to the dynamic, ever evolving inner regions of Mira's dusty circumstellar environment.

This technique has the potential to obtain polychromatic stellar images from a space platform at similar angular resolutions to terrestrial long-baseline interferometry.
It can be applied to both archived and new observations (until \emph{Cassini's} planned demise in 2017) providing they have appropriate occultation geometry.
Its application to other stellar systems could potentially yield reveal additional spatial information about circumstellar environments.

\section*{Acknowledgements}
The authors would like to thank Anthony Alexander, Dr Anne Rogerson, Nick Olson, and Benjamin Pope for their helpful advice on Greek usage in the title.

\bibliographystyle{mn2e} 
\bibliography{library}

\bsp

\label{lastpage}

\end{document}